\documentclass[12pt]{article}

\usepackage{amsfonts, amsmath}
\usepackage{bbm}
\usepackage{epsf}

\newcommand{\bmat}{\left(\begin{array}}
\newcommand{\emat}{\end{array}\right)}

\def\a{\alpha}

\def\-{\hphantom{-}}
\def\ov{\overline}
\def\s2{\frac{1}{\sqrt2}}

\def\oh{\frac{1}{2}}
\def\beq{\begin{equation}}
\def\eeq{\end{equation}}
\def\beqa{\begin{eqnarray}}
\def\eeqa{\end{eqnarray}}
\def\D{{\rm D}}

\def\pp{{\mathbb P}}
\def\complexp{{\mathbb C}}

\def\cd{{\mathcal D}}

\def\cn{{\mathcal N}}

\def\co{{\mathcal O}}

\def\neq#1{\mbox{$\cn$=#1}}
\def\Dsl{\,\raise.15ex\hbox{/}\mkern-13.5mu D} 
\def\r#1{\mbox{{\bf #1}}}

\def\r#1{{\bf #1}}
\def\br#1{{\bf \overline{#1}}}

\def\RR{{\mathbf{R}}}
\def\rr{{\mathbf{r}}}
\begin{document}
\pagestyle{plain}

\makeatletter
\@addtoreset{equation}{section}
\makeatother
\renewcommand{\theequation}{\thesection.\arabic{equation}}
\pagestyle{empty}
\rightline{ IFT-UAM/CSIC-08-75}
\vspace{0.5cm}
\begin{center}
\LARGE{ Yukawa  Structure from $U(1)$  Fluxes in \\
  F-theory  Grand  Unification 
\\[10mm]}
\large{A. Font$^1$ and L.E. Ib\'a\~nez$^2$ \\[6mm]}
\small{
${}^1$  Departamento de F\'{\i}sica, Centro de F\'{\i}sica Te\'orica y Computacional \\[-0.3em]
 Facultad de Ciencias, Universidad Central de Venezuela\\[-0.3em]
 A.P. 20513, Caracas 1020-A, Venezuela\\[2mm] 
${}^2$ Departamento de F\'{\i}sica Te\'orica 
and Instituto de F\'{\i}sica Te\'orica UAM-CSIC,\\[-0.3em]
Universidad Aut\'onoma de Madrid,
Cantoblanco, 28049 Madrid, Spain 
\\[8mm]} 
\small{\bf Abstract} \\[5mm]
\end{center}
\begin{center}
\begin{minipage}[h]{14.0cm} 
In F-theory GUT constructions Yukawa couplings 
necessarily take
place at the intersection of three matter curves.
For generic geometric configurations this 
gives rise to problematic Yukawa couplings 
unable to reproduce the observed hierarchies.
We point out that if the $U(1)_{B-L}$/$U(1)_Y$
flux breaking the $SO(10)/SU(5)$ GUT symmetry is 
allowed to go through pairs of matter curves
with the same GUT representation, the quark/lepton 
content  is redistributed in such a way that all 
quark and leptons are allowed to have hierarchical
Yukawas. This reshuffling of
fermions is quite unique and is
 particularly elegant  for the case of three
generations and $SO(10)$.
Specific local F-theory models  with
$SO(10)$ or $SU(5)$ living on a del Pezzo  surface with appropriate
bundles and just  the massless content of the MSSM are
described. We point out that the smallness 
of the 3rd generation quark mixing predicted by this scheme 
(together with gauge coupling unification) could constitute a first hint
of an underlying F-theory grand unification.

\end{minipage}
\end{center}
\newpage
\setcounter{page}{1}
\pagestyle{plain}
\renewcommand{\thefootnote}{\arabic{footnote}}
\setcounter{footnote}{0}




\section{Introduction}

In the search for string compactifications with low-energy physics as close as
possible to observations  two approaches seem feasible.
In a top-bottom approach one  starts from some string compactification 
which is fully consistent globally (e.g. with global RR-tadpole
cancellation in the Type II case) and after a process of symmetry breaking 
one obtains a low-energy spectrum  close to the SM (or the MSSM). 
On the other hand, in a bottom-up approach \cite{aiqu} one considers local configurations of lower 
dimensional $\D p$-branes, $p\leq 7$, which are localized on some region of the
compact dimensions and reproduce SM physics. In this second case one does not care 
about the global aspects of the compactification and assumes that eventually the
configuration may be embedded inside a fully consistent global model.
This second bottom-up approach is not available in heterotic or Type I
string compactifications since the SM fields would then live in the full 
six extra dimensions.

In principle one would say that having a globally consistent compactification
would be more satisfactory. However, local configurations of branes may be
more efficient in trying to identify promising classes of string vacua, independently
of the details of the global theory. Eventually, if one successful bottom-up 
configuration is found one can then look for ways to embed it in a globally consistent model.
In a stronger version of a local construction, the local SM or GUT physics is required to
decouple from the gravitational sector in the infinite volume limit. This is the case of 
models derived from $\D 3$-branes at singularities \cite{aiqu,bjl,cgqu,vw,cmq} in which 
the SM physics only depends on the local geometry  around the singularity \cite{aiqu}.
In this situation one can cleanly disconnect the SM physics from the gravitational sector
in such a way that local model building has an specific sense, since 
SM physics may indeed be separated from the rest of the compactification.

Another realization of the bottom-up philosophy has been recently 
considered in the context of local F-theory \cite{ftheory} grand unified models \cite{dw,bhv1,bhv2}.
As particularly emphasized in \cite{bhv1,bhv2}, if the F-theory 7-branes 
containing a GUT group wrap certain classes of complex two-dimensional surfaces $S$ 
whose volume is contractible to zero (del Pezzo surfaces), one can take the overall infinite
volume limit ($M_p\rightarrow \infty $) and get again the SM/GUT physics decoupled from
the gravitational sector. The main advantage of a F-theory GUT version of the
bottom-up approach is that gauge coupling unification is naturally guaranteed.
In the context of the MSSM the gauge coupling unification prediction
\cite{gcu} is in very good agreement with experiment and it seems sensible 
trying to incorporate it when  searching  for a realistic string vacuum. 
In an independent development it has been shown \cite{aci} that if MSSM matter is localized 
on 7-brane intersections and the source of SUSY breaking is modulus mediation, the emerging pattern 
of SUSY-breaking soft terms at tree level is consistent with all current 
experimental constraints. It was also shown that in order to obtain these
successful results it was important that in the physical Yukawa couplings 
driving electroweak symmetry breaking all the fields involved should come from
intersecting 7-branes. Interestingly enough this  is the structure that 
appears in the F-theory GUT's constructed in \cite{bhv2} in which indeed
non-vanishing Yukawa couplings appear only among fields living
at intersecting 7-branes.

In the class of local F-theory GUT's constructed in \cite{bhv1,bhv2} 
(see also \cite{tw,httwy,hmssv,hv1,hv2,mss1,mss2,w,dw2} for 
more recent work) the MSSM fields reside at certain matter curves
corresponding to Riemann surfaces on the complex 2-fold $S$  
where the GUT symmetry lives. At those curves the singularity associated to the 
GUT symmetry ($SU(5)$ or $SO(10)$) is enhanced and at points at  which three
matter curves  intersect there is a further enhancing of the singularity.
These triple intersections of two quark/lepton matter curves 
with a Higgs curve enable Yukawa couplings to appear.  However, there is 
a feature of the Yukawa couplings which is quite unattractive. 
For a Yukawa coupling to exist, the two matter curves associated 
to quarks/leptons must be different. Thus, for example, 
in an $SU(5)$ F-theory GUT of this class a coupling 
$\r{10}_i\times \r{10}_j\times \r{5}_H$ is only non-vanishing when $i\not=j$.
As emphasized in \cite{bhv2} such kind of U-quark mass matrix
is unable to accommodate the hierarchical structure observed 
experimentally. In \cite{bhv2} a solution to this
{\it F-theory GUT's Yukawa problem} was proposed in which it is assumed 
that the $\r{10}$s (or the $\r{16}$'s in $SO(10)$ GUT's) have 
self-pinching, a self-intersecting geometry in which the corresponding matter
curve intersects itself  on the surface $S$.

In this paper we point out that the F-theory Yukawa problem
is naturally solved by slightly generalizing the conditions
on the $U(1)$ fluxes breaking the GUT symmetry down to the SM. 
Indeed, in order to avoid that different MSSM particles  
could have different multiplicities it is assumed in \cite{bhv2}
that the $U(1)$ flux through all matter curves $\Sigma_i$ 
containing quark and leptons vanish,
\beq
\int_{\Sigma_i} \ F_{U(1)} \ =\ 0  \ .
\eeq
Our main point is that this condition is too strong. 
It is enough to ask that if we have two matter curves 
$\Sigma_1$ and $\Sigma_2$ corresponding to the same GUT representation,
in order to keep equal the number of generations for all
quarks and leptons it is enough to request that
this is true on average, i.e.
\beq
\int_{\Sigma_1} \ F_{U(1)} \ +\ \int_{\Sigma_2}   \ F_{U(1)} \ =\ 0 \ .
\eeq
If this is the case, one  finds  that the left-handed and right-handed
quarks and leptons are redistributed unequally among the
two matter curves $\Sigma_1$ and $\Sigma_2$ in such a way that 
all quarks and leptons are allowed to get non-vanishing Yukawa 
couplings in a way which allows for a hierarchical structure.
Moreover, as a direct byproduct it follows that 
the mixing of the third generation quarks with the first two generations 
is suppressed. On the contrary, mixing among leptons is unconstrained.
These two facts are in good qualitative agreement with
experiment. We find this fact quite encouraging since 
we were only looking for a way to get Yukawa couplings 
for all quarks and leptons and not looking for any particular texture.
This works both for  $SO(10)$ and $SU(5)$ F-theory GUT's, although for
$SO(10)$ this reshuffling of quarks and leptons is particularly
simple and  symmetric. In the $SO(10)$ case it is the flux
of the $U(1)_{B-L}$ in the bulk which breaks the symmetry down to the
left-right symmetric model and hence a further step
of symmetry breaking down to the SM is needed. In the case of
$SU(5)$ it is the hypercharge flux which breaks the bulk symmetry down to the SM.
However the matter curves will feel $U(1)_{B-L}$ flux which again 
reshuffles the fermion spectrum in the matter curves associated to $\r{10}$s and 
$\br{5}$s.
Although the number of quark/lepton generations is not fixed in these 
local F-theory constructions, it is intriguing that three generations
is somewhat special. Indeed one can see that three 
is the minimum number of generations consistent with having a purely 
chiral quark/lepton spectrum (i.e. without vector-like 
fermion copies). 

The structure of this paper is as follows. After reviewing the
F-theory GUT constructions of \cite{bhv1,bhv2}, we 
discuss in section 3 the {\it F-theory GUT Yukawa problem} and its 
solution in terms of the reshuffling of fermions induced 
by $U(1)$ fluxes in general terms. In section 4 we construct an specific 
local F-theory $SO(10)$ GUT with the symmetry broken by 
$U(1)_{B-L}$ flux. A brief discussion of some phenomenological
aspects is given. An $SU(5)$ local GUT is presented in section 5.
Some final comments are left for section 6.
A brief compendium on del Pezzo surfaces and other mathematical
results are collected in an appendix.

\section{F-theory Grand Unification}

In this section we briefly review the basic formalism. We mostly follow \cite{bhv1,bhv2}
where the reader can find a more complete discussion (see also \cite{dw,w}). 
Some  details about del Pezzo surfaces and other useful results are 
summarized in the appendix.

In F-theory GUT's the gauge theory arises from 7-branes wrapping a compact surface $S$
of complex codimension one in the base of an elliptically-fibered Calabi-Yau fourfold. 
The gauge group $G_S$ depends on the singularity type of the elliptic fiber. This gauge group 
can be broken by a background  in a subgroup $H_S \subset G_S$. We will consider $H_S=U(1)$.
\neq1 supersymmetry requires this $U(1)$ line bundle $L$ to satisfy certain conditions
stated in \cite{bhv1}. 

We will take $S$ to be a del Pezzo surface. As stressed in \cite{bhv1, bhv2}, in this case the resulting 
gauge theory on the 7-branes decouples from gravity in the bulk. In consequence the resulting GUT is more
constrained. In particular, the allowed supersymmetric line bundles are completely characterized as 
reviewed in the appendix. In this framework it is possible to extract definite physical features
of F-theory GUT's.

The $U(1)$ background breaks the gauge group $G_S$ to the commutant $\Gamma_S \times U(1)$.
The adjoint representation of $G_S$ decomposes into a direct sum of irreducible representations
of $\Gamma_S \times U(1)$, each generically labelled as $(\rr,q)$.  
When $S$ is a del Pezzo surface, the number of multiplets from the adjoint with $U(1)$ charge 
$q$, and transforming under $\rr$ of $\Gamma_S$, is given by 
\beq
N(\rr, q) = h^1(S, L^q)=-\big[ 1 + \oh c_1(L^q) \cdot c_1(L^q)\big]  \ ,
\label{indexq}
\eeq   
where $h^1(S, L^q)$ is the dimension of the corresponding cohomology group and $c_1(L^q)$
is the first Chern class of the bundle $L^q$.
Here we have already used that the admissible line bundles on a del Pezzo surface must satisfy
$c_1(S)\cdot c_1(L)=0$. 
Observe that the multiplicity does not depend on the sign of $q$. 
Thus, $N(\rr, q)=N(\bar \rr, -q)$, and only vector-like matter will result.

Charged multiplets with definite chirality will never originate from the 
adjoint in the breaking of $G_S$ when $S$ is a 
del Pezzo surface. Fortunately, in general there is another way to obtain charged matter in F-theory
GUTs. Charged multiplets also arise by introducing non-compact surfaces $S_i^\prime$
asociated to extra 7-branes, each intersecting 
$S$ at a Riemann surface $\Sigma_i$ where the fields are localized. At each matter curve $\Sigma_i$ 
the singularity type is enhanced to group $G_{\Sigma_i}$ of rank at least one higher. 
The gauge group $G_S$ is further enhanced to $G_p$ of rank at least two higher at points in $S$ 
where three matter curves intersect. For instance, when $G_p$ has rank two higher in practice
$G_p \supset G_S \times U(1)_a \times U(1)_b$, and from
the decomposition of the adjoint of $G_p$ one can identify the intersecting matter curves.
The multiplets that materialize at each intersection descend from the adjoint of $G_{\Sigma_i}$
and their degeneracy can be computed as reviewed shortly.  

To be more concrete, we consider an example with $G_S=SO(10)$ and $G_p=E_7$. We need the
decomposition of the adjoint $\r{133}$ given by
\beqa
E_7 & \supset &  SO(10) \times U(1)_a \times U(1)_b \label{rep133} \\[2mm]
\r{133} & = & {\rm Adjoints} +  [(\r{10},2,0) + (\r{16},-1,1)  + (\r{16},-1,-1) + (\r1,0, 2)  
+ {\rm c.c.}] \ . \nonumber 
\eeqa
The last entries are the $U(1)$ charges.
The $U(1)$ generators will be denoted $Q_a$ and $Q_b$.
The main point is that for each distinct charged representation that shows up in the adjoint 
branching there is a matter curve. In this example there are two distinct $\r{16}$'s and a Higgs $\r{10}$.
Thus, there will be quark-lepton curves $\Sigma_1$ and $\Sigma_2$
associated each to a $\r{16}_i$, and a Higgs curve $\Sigma_H$ that hosts the $\r{10}_H$.
Since the curves are required to intersect, there will be a coupling 
$\r{16}_1\times \r{16}_2 \times\r{10}_H$, which is indeed allowed by gauge invariance.

On each matter curve $G_S$ is enhanced to a group of rank at least one higher. In the
$G_S=SO(10)$ and $G_p=E_7$ example, the singularity is enhanced to $G_{\Sigma_i}=E_6$
on $\Sigma_i$, $i=1,2$. Decomposing the adjoint of $E_6$ under $SO(10) \times U(1)_i^\prime$
includes a $\r{16}_i$. To determine the $U(1)_i^\prime$ generators, notice that (\ref{rep133})
shows that along the directions $(Q_a-Q_b)=0$ and $(Q_a+Q_b)=0$ there appear 32 additional neutral states
and $SO(10)$ is indeed enhanced to $E_6$.    
Choosing a convenient normalization we then have  
\beq
Q^\prime_1 = -\frac12 (Q_a-Q_b) \quad ; \quad Q^\prime_2 = -\frac12 (Q_a+Q_b) \ .
\label{q12def}
\eeq
In this way there will be a $\r{16}_1$ with charge $q^\prime_1=1$, and a $\r{16}_2$ with 
$q^\prime_2=1$. Concerning the Higgs curve, the singularity must be enhanced to $G_{\Sigma_H}=SO(12)$.
Decomposing the adjoint of $SO(12)$ under $SO(10) \times U(1)_H^\prime$ yields the desired $\r{10}_H$.
{}From (\ref{rep133}) we see that $SO(10) \to SO(12)$ along $Q_a=0$.
We choose $Q^\prime_H=Q_a/2$ and $q^\prime_H=1$.

Following \cite{bhv1, bhv2} we now discuss how to compute the degeneracy of the multiplets 
living on a matter curve $\Sigma$ at the intersection of $S$ and some $S^\prime$. 
On $\Sigma$ the GUT group $G_S$ is enhanced to $G_\Sigma \supset G_S \times G_{S^\prime}$. We consider
$G_{S^\prime}=U(1)^\prime$. In turn $G_S$ is broken to $\Gamma_S \times U(1)$ by a background
in $H_S=U(1)$. Similarly, $G_{S^\prime}$ is broken by a background $H_{S^\prime}$ that is 
taken to be $U(1)^\prime$.  The two line bundles are denoted $L$ and  $L^\prime$ respectively.
The adjoint of $G_\Sigma$ decomposes under $\Gamma_S \times U(1) \times U(1)^\prime$
into a direct sum of irreducible representations that can be labelled $(\rr, q, q^\prime)$. 
The degeneracy of a multiplet transforming into such a representation is given by  
\beq
N(\rr, q, q^\prime) = h^0(\Sigma, K_{\Sigma}^{1/2}\otimes L^q_{\Sigma}
\otimes L^{\prime q^\prime}_{\Sigma}) \ ,
\label{multqq}
\eeq
where $L_\Sigma \equiv L |_{\Sigma}$ and $L^\prime_\Sigma \equiv L^\prime|_{\Sigma}$ are the restrictions of
the line bundles $L$ and $L^\prime$ to $\Sigma$. Here $K_\Sigma$ is the canonical line bundle of $\Sigma$.
We will mostly consider the cases where $\Sigma$ has genus zero and $ K_{\Sigma}^{1/2}=\co_\Sigma(-1)$,
or $\Sigma$ has genus one and $K_\Sigma$ is trivial.
For the complex conjugate it follows that
\beq
N(\ov{\rr}, -q, -q^\prime) = h^1(\Sigma, K_{\Sigma}^{1/2}\otimes L^q_{\Sigma}
\otimes L^{\prime q^\prime}_{\Sigma}) \ .
\label{multqqc}
\eeq
Furthermore, taking into account that $L$ and $L^\prime$ are line bundles, the net multiplicity turns out to be
\beq
N(\rr,q,q^\prime)- N(\ov{\rr},-q,-q^\prime)=(1-g) + 
c_1(K_{\Sigma}^{1/2}\otimes L^q_{\Sigma} \otimes L^{\prime q^\prime}_{\Sigma}) \ ,
\label{indexqq}
\eeq
where $g$ is the genus of $\Sigma$. Finally, we remark that the $U(1)^\prime$ 
decouples because the non-compact surface $S^\prime$ has formally infinite volume. Then, after computing the 
degeneracies we can drop the label $q^\prime$. 

In the finite volume case the extra $U(1)$ symmetries are anomalous
and their gauge bosons get generically massive by combining with RR-fields 
of the full compactification. Note
that, as pointed out in \cite{imr,aiqu},  anomaly free 
$U(1)$'s like hypercharge or $U(1)_{B-L}$ may also become massive
by combining with RR fields. There are  conditions under which this may be 
avoided \cite{imr,aiqu} (see also \cite{bmmvw}) and in the F-theory
GUT's this can also  be implemented \cite{bhv2}.
This is not the case in heterotic compactifications in which 
setting flux along some $U(1)$ direction necessarily gives mass
to the corresponding gauge boson \cite{bhv2} (see also \cite{tw}).

In the appendix we will further discuss the quantities involved in computing the multiplicities
and will collect some useful results to this purpose.

\section{$\pmb{U(1)}$ Fluxes and fermion splitting}

\begin{figure}
\epsfysize=8cm
\begin{center}
\leavevmode
\epsffile{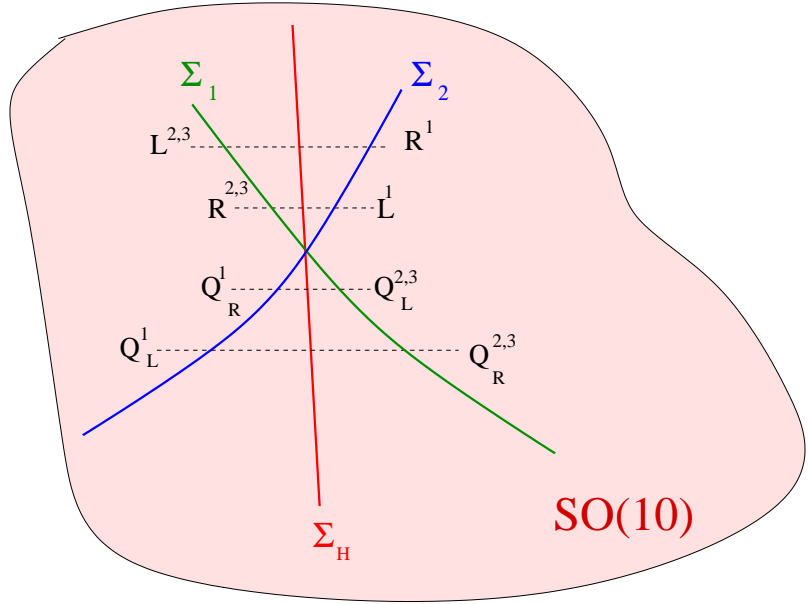}
\end{center}
\caption{\small Triple intersection of two curves $\Sigma_1$ and $\Sigma_2$,
with two and one generations of $SO(10)$ without fluxes going through them.
Dashed lines represent fermions  linked by a Yukawa coupling.
Most Yukawa couplings are forbidden.}
\label{calvicie}
\end{figure}

As mentioned before, there is a generic problem
for F-theory GUT's from the Yukawa coupling structure. Indeed, 
in these models Yukawa couplings come from the intersection of three curves 
inside the surface $S$. These curves generically correspond to different matter
fields and the coupling constants have the structure \cite{bhv2}
\beq
Y_{ij} \ =\ \sum_p \ \Psi_R(p)^i\Psi_L(p)^j\Psi_H(p)
\label{Yukawas}
\eeq
where $\Psi_{R,L,H}(p)$ are the internal wave functions of the  
 left-handed, right-handed and Higgs matter fields evaluated at the
intersection points $p$, which may be generically multiple. 
In $SO(10)$ GUT's the Yukawa couplings are of the form
$\r{16}_i \times \r{16}_j \times \r{10}_H$ and the
fact that the curves involved must be different  means that for $i=j$ 
those Yukawa couplings vanish, i.e. the
structure of Yukawa couplings is of the form
\beq
Y_{q,l}\ \simeq \
\left(
\begin{array}{ccc}
0 & x & x \\
x & 0 & x \\
x & x & 0
\end{array}
\right)
\label{yukft}
\eeq
with $x$ representing generically non-vanishing 
entries.
This is illustrated in figure \ref{calvicie} for a particular 
$SO(10)$ example with  two  matter curves $\Sigma_1$ and $\Sigma_2$ having two
and one generations respectively. From (\ref{yukft}) one can then easily see that 
there is no regime in which one of the eigenvalues of the
matrix is large with the other two small, as required to 
explain the data. Rather if one of the eigenvalues is small,
the heavier ones are necessarily of the same order of magnitude. 
Something analogous happens in $SU(5)$ for the
U-quark masses which come from couplings with the structure 
$\r{10}_i \times \r{10}_j \times \r{5}_H$. Here again it is not possible 
to obtain one large eigenvalue (top quark) with the other two
much smaller.

A possible solution to this problem advocated in \cite{bhv1,bhv2} 
consists of assuming that the matter curves where the fermions in
problematic  representations reside ($\r{16}$s in $SO(10)$, 
$\r{10}$s in $SU(5)$) have a {\it self-intersecting}
geometry, the corresponding curve intersects itself inside $S$.
 This in general provides for non-vanishing 
diagonal entries in eq.(\ref{yukft}) which could then be
compatible with a hierarchical structure.
Although this could certainly be a solution to the
F-theory GUT Yukawa problem, we would like to argue in the present article 
that there is quite a natural and economical solution which
does not require the assumption of any intricate geometry 
for the matter curves but rather relies on the gauge structure of the GUT.

Given that Yukawa terms couple right-handed and  left-handed fermions to
a Higgs field, the idea of the natural solution is to have a matter curve $\Sigma_R$ with all 
right-handed fermions, another one $\Sigma_L$ with all left-handed fermions and both
intersecting a Higgs matter curve $\Sigma_H$. In this way 
all Yukawa couplings would be in principle allowed, there would be
no self-intersections required. This would be for example the case 
in perturbative Type IIB orientifolds with  e.g. a 
$SU(4)\times SU(2)_L\times SU(2)_R$ structure in which 
left- and right-handed quarks may be localized  at 
different D7-brane intersections and transform as 
$(\r{4},\r{2}_L,\r1)$ and $(\br{4}, \r1, \r{2}_R)$ respectively.
However our philosophy is to build GUT models which incorporate gauge coupling unification in a
natural way. Unfortunately in GUT's like $SO(10)$ or $SU(5)$ this structure is not available
since left- and right-handed fermions
sit in the {\it same}  GUT representations. 
However, there is a logical way out in F-theory GUTs.
Imagine we start with two matter curves $\Sigma_1$ and
$\Sigma_2$,  each containing matter in the {\it same}
GUT representation $\RR$, e.g. $\r{16}$ in $SO(10)$, or $\r{10}$ 
in $SU(5)$. The same $U(1)$ flux which breaks the GUT symmetry down to the
SM (or some extension) could perhaps be allowed to go through 
both  matter curves and split the fields so that
one curve contains right-handed fields and the other
left-handed fields. Then again all Yukawa couplings would 
be allowed. Indeed we will see that $U(1)_{B-L}$ fluxes in
$SO(10)$ or hypercharge flux in $SU(5)$ can under
certain conditions reshuffle the matter content in curves
$\Sigma_1$ and $\Sigma_2$ in such a way 
that Yukawa couplings for all quarks and leptons 
may appear. In fact there will not be  
purely right-handed $\Sigma_R$ or left-handed $\Sigma_L$ 
curves but the net effect will be quite analogous to
that structure with an appropriate {\it flipping}
of left by right in the curves.

In the models constructed in \cite{bhv1,bhv2} the net flux through each matter curve containing 
quarks and leptons is assumed to be zero. Indeed, threading flux through the quark-lepton curves 
seems risky. Take for example the $SU(5)$ case. If hypercharge flux is 
added, since the SM fields have different hypercharges,
one would expect different multiplicity for each SM
field, nothing we would like to have.
However this is not true in general. As we will see, if we have two matter curves, 
$\Sigma_1$ and $\Sigma_2$, giving rise to the same GUT group representation, there are simple
conditions under which a $U(1)$ flux may be allowed to go through the two curves 
in such a way that the net multiplicity for each multiplet is equal, despite all of them
having different charges. 

Let us be more specific. Consider two matter curves $\Sigma_1$ and $\Sigma_2$ which we will take for
definiteness to have genus $g=0$. Each curve contains matter transforming in the same GUT 
representation  $\RR_1=\RR_2$. We are adding fluxes through both curves along a \mbox{$U(1)\subset G_S$}, 
with $G_S$ the GUT group living on the surface $S$. In addition these representations have charges 
$q_1^\prime$ and  $q_2^\prime$ under groups $U(1)_1^\prime$ and $U(1)_2^\prime$ 
corresponding to the intersecting surfaces $S_1^\prime$ and $S_2^\prime$.
Once the $U(1)$ flux is added the GUT symmetry is broken to $\Gamma_S \times U(1)$ and the
two multiplets further decompose as 
\beq
\RR_1 \ =\ \bigoplus_\a (\rr_\a, q_\a, q_1^\prime) \quad ;\quad 
\RR_2 \ =\ \bigoplus_\a (\rr_\a, q_\a, q_2^\prime) \ ,
\label{dr1r2}  
\eeq
where the $\rr_\a$ are irreducible representations of $\Gamma_S$.

We now assume that the restrictions of the $U(1)$ bundle $L$ 
to the two curves have the forms $L_{\Sigma_1}=\co_{\Sigma_1}(1)^{1/n}$ and 
$L_{\Sigma_2}=\co_{\Sigma_2}(-1)^{1/n}$ respectively. We have allowed for possible fractional bundles,
i.e. $n$= 5 for $SU(5)$, and $n=4,2$ for $SO(10)$ \cite{bhv2}. The main point here is that 
$L_{\Sigma_1}$ and $L_{\Sigma_2}$ have opposite instanton numbers. 
Instead of insisting that the fluxes through each individual curve
vanish, we rather impose that the flux vanishes on average, namely 
\beq
\int_{\Sigma_1} \ F_{U(1)} \ +\ \int_{\Sigma_2}   \ F_{U(1)} \ =\ 0 \ .
\eeq
There will also be the restrictions of the gauge
bundles on $S_1'$ and $S_2'$ which we take to be of the general form
$L^\prime_{\Sigma_1}=\co_{\Sigma_1}(n_1)^{1/n}$ and $L^\prime_{\Sigma_2}=\co_{\Sigma_2}(n_2)^{1/n}$ 
with $n_1$ and $n_2$ appropriately chosen integers.
Under these conditions the multiplicity of each representation $(\rr_\a,q_\a)$ due to each curve will be
(we are assuming $g=0$ for both curves) 
\beqa
\Sigma_1  &:&  N_1(\rr_\a,q_\a)\ =\ h^0(\Sigma_1, \co_{\Sigma_1}(-1 + 
\frac {1}{n} [ q_\a + n_1 q_1^\prime]) \ =\ \frac {1}{n} (q_\a + n_1 q_1^\prime) \label{idea} \\[2mm]
\Sigma_2  &:&  N_2(\rr_\a, q_\a)\ =\ h^0(\Sigma_2, \co_{\Sigma_2}(-1 +
\frac {1}{n} [ -q_\a + n_2 q_2^\prime]) \ =\ \frac {1}{n} (-q_\a + n_2 q_2^\prime) \nonumber
\eeqa
so that the net multiplicity of each representation  $(\rr_\a,q_\a)$  is given by
\beq
N (\rr_\a,q_\a) \ =\ \frac {1}{n} \ (n_1q_1^\prime\ +\ n_2q_2^\prime)
\eeq
\label{numsem}
Note that this multiplicity is {\it independent of the value of the charge $q_\a$} 
of each of the multiplets $\rr_\a$ inside the GUT representation $\RR$. Thus, the same
multiplicity for all different components can result irrespective of their charges. 
Note also that if  $|n_1 q_1^\prime|$  or $|n_2q_2^\prime|$  are smaller than some  $|q_\a|$  the
spectrum will then include additional vector-like multiplets of charge $q_\a$.
The crucial conclusion is that even though the net number of each $(\rr_\a, q_\a)$ representation 
is the same, they will be distributed unequally between the two matter curves $\Sigma_1$ and $\Sigma_2$.

Note that this may also be understood as a single reducible matter curve such that
the net flux through it  is zero. Furthermore it may trivially extended to the case of
more than two matter curves in which the net flux vanishes.

In the following we apply this simple idea to specific GUT models
with two matter curves $\Sigma_1$ and $\Sigma_2$ 
and see how the Yukawa problem is then generically solved.

\section{A $\pmb{B}{\bf -}\pmb{L}$ fluxed  $\pmb{SO(10)}$ model}

\begin{figure}
\epsfysize=8cm
\begin{center}
\leavevmode
\epsffile{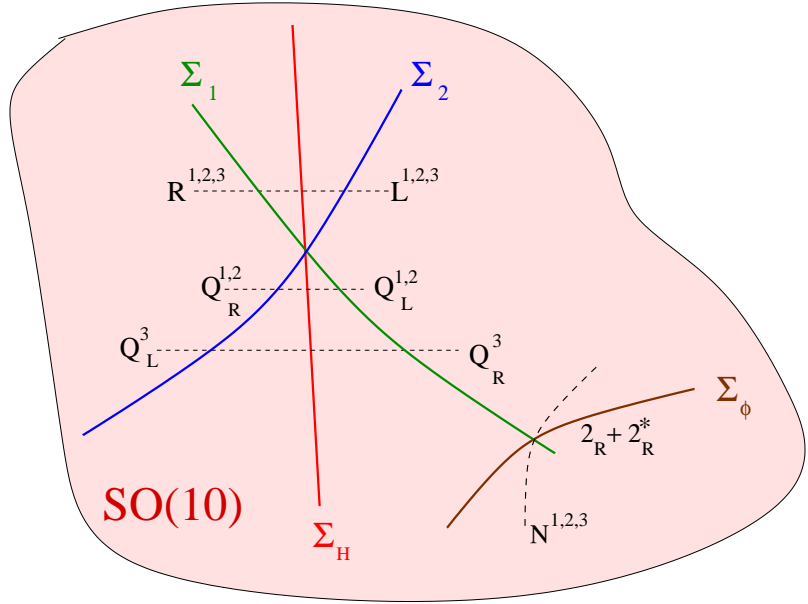}
\end{center}
\caption{\small Sketch of the structure of the $B-L$ fluxed $SO(10)$ model.
Dashed lines represent fermions  linked by a Yukawa coupling.
Yukawa couplings for all fermions are allowed.}
\label{so10quiver}
\end{figure}

Let us first consider the case of an $SO(10)$ GUT which is broken
by $U(1)_{B-L}$ flux down to a left-right symmetric extension of
the SM. A possible choice of curves and bundles is shown in table
\ref{so10}
(see the appendix for notation). A sketch of the matter  curves involved is depicted
in figure \ref{so10quiver}.

\begin{table}[htb] \footnotesize
\renewcommand{\arraystretch}{1.25}
\begin{center}
\begin{tabular}{|c|c|c|c|c|c|}
\hline
 Multiplet &  Curve  &  Class  & $g_\Sigma $ & $ L_\Sigma $ & $L'_\Sigma $ \\
\hline\hline
$ \r{16}_1$  & $\Sigma_1$ &  $H-E_1-E_3$ &    0   &  $\co_{\Sigma_1}(1)^{1/2}$ & $\co_{\Sigma_1}(3)^{1/2}$  \\
\hline
$\r{16}_2$  & $\Sigma_2$ &  $H-E_2-E_4$ &  0   &  $\co_{\Sigma_2}(-1)^{1/2}$ &  $\co_{\Sigma_2}(3)^{1/2}$  \\
\hline
$\r{10}$  & $\Sigma_{H}$ &  $-K_S$ &  1    &  $\co_{\Sigma_H}(p_1-p_2)^{1/2} $ &  $\co_{\Sigma_H}$   \\
\hline
$(\r{16}+\br{16})$   & $\Sigma_{\phi}$ &  $  3H-E_1 -E_2$ &  1    &  $\co_{\Sigma_\phi}(p_3-p_4)^{1/2}$ & 
$\co_{\Sigma_\phi}(p_3-p_4)^{-3/2}$ \\  
\hline 
\end{tabular}
\end{center} \caption{\small  Curves and bundles of the $SO(10)$ model with $U(1)_{B-L}$ flux
and \mbox{$L=\co_S(E_1-E_2)^{1/2}$}.}
\label{so10}
\end{table}

We have F-theory 7-branes wrapping a del Pezzo surface $S$ with a singularity corresponding
to a $SO(10)$ gauge symmetry. Adding $U(1)_{B-L}$ flux breaks the symmetry 
and the adjoint decomposes as
\beqa
SO(10) & \supset  &  SU(3)\times SU(2)_R \times SU(2)_L \times U(1)_{B-L} \label{branchingubl} \\[2mm] 
\r{45} & =  & \ {\rm Adjoints} \ + \ (\br{3}, \r1,\r1,-4) + (\r3, \r1,\r1,4) + 
(\br{3},\r2,\r2,2) + (\r3,\r2,\r2,-2) 
\nonumber
\eeqa
where the last entry is the $B{\rm -}L$ charge. Upon symmetry breaking the $SO(10)$ adjoint gives rise
to fields transforming under the unbroken gauge group according to this decomposition. 

In general in the presence of flux there may be massless exotics from the adjoint transforming
as $(\br{3},\r2,\r2,2)$, $(\r3,\r1,\r1,4)$ and/or their conjugates.
As discussed in \cite{bhv2}, in the case of $SU(5)$,  exotics disappear if some appropriate
powers of the $U(1)$ bundle correspond to a $dP_N$ divisor associated to $E_8$ roots. 
In $SO(10)$ models one cannot get rid of all exotics \cite{bhv2}, but one can still remove
most of them. To this end we make the choice
\beq
L=\co_S(E_1-E_2)^{1/2} \ ,
\label{lbml}
\eeq 
where the $E_i$ are exceptional divisors. We can now compute the multiplicity
of the states descending from the adjoint using (\ref{indexq}). We obtain
\beq
      N(\br{3},\r2,\r2,2)  \  = \ h^1(S, L^2) \ = \ 0 \quad  ; \quad
                 N(\r3,\r2,\r2,-2) \ = \ h^1(S, L^{-2})\ =\ 0
\eeq
because $c_1(L^2)\cdot c_1(L^2)=-2$, reflecting that $L^2$ indeed corresponds to an $E_8$ root. 
On the other hand,
\beq
                N(\br{3},\r1,\r1,-4) \   =  \ h^1(S, L^{-4})
= \ -\big[ 1   +  \oh c_1(L^{-4})^2\big] \  = \ 3
\eeq
and the same result for $N(\r3,\r1,\r1,4) = h^1(S, L^4)$.
So we will have three sets of vector-like chiral multiplets
\beq
                        3[(\br{3},\r1,\r1,-4) + (\r3,\r1,\r1,4)]
 \eeq
In fact, after further symmetry breaking down to the SM
these fields  have the quantum numbers of vector-like right-handed D-quarks.
We will argue later that these states are expected to become massive after
the  symmetry breaking process from $SU(3)\times SU(2)_R\times SU(2)_L\times U(1)_{B-L}$ down to the SM.

We have an $SO(10)$  gauge group in our $S$ surface.  At some complex curves  $\Sigma_i$
the rank of the singularity is enhanced  so that matter transforming as both $\r{16}$ and $\r{10}$
does appear. In particular,  $\r{16}$s arise from curves in which the singularity is
enhanced to $E_6$ whereas $\r{10}$s appear at curves where the symmetry is enhanced to $SO(12)$.
These curves will intersect at points of double enhancing up to $E_7$. Recalling the decomposition of
the adjoint of $E_7$ under $SO(10)\times  U(1)_a \times U(1)_b$ displayed in (\ref{rep133}) we see
that the coupling $(\r{16},-1,1)\times (\r{16},-1,-1) \times (\r{10},2,0)$ is allowed   
so that in principle one can engineer three curves $\Sigma_i$ , $i=1,2$, and $\Sigma_{H}$, associated to 
each representation and giving rise to Yukawa couplings. In particular, we have two types of
matter curves $\Sigma_1$ and $\Sigma_2$ supporting $\r{16}$s that we can use
for the fermion reshuffling mechanism described in the previous section.

We have chosen the matter and Higgs curves as  indicated in table \ref{so10}.
They verify the condition that $\Sigma_1$ and $\Sigma_2$ have non-trivial (and positive) intersection 
among themselves as well as with the Higgs curve $\Sigma_{H}$, so that Yukawa couplings are possible.
The matter curves are also chosen so that the restrictions of the $U(1)_{B-L}$ line bundle 
$L=\co_S(E_1-E_2)^{1/2}$ to the $\Sigma_i$ have opposite instanton numbers. Since 
$(E_1-E_2)\cdot\Sigma_1=1$ and $(E_1-E_2)\cdot \Sigma_2=-1$ we find that
\beqa
L_{\Sigma_1} &=& \co_{\Sigma_1}(1)^{1/2} \ , \nonumber \\
L_{\Sigma_2} &=& \co_{\Sigma_2}(-1)^{1/2} \ , \label{instcond} 
\eeqa
as required in order to have $\int_{\Sigma_1}F_{U(1)_{B-L}}+\int_{\Sigma_2}F_{U(1)_{B-L}}=0$.

The curve $\Sigma_H$ supporting the Higgses that couple to matter fermions is selected to have genus one.
Doublet-triplet splitting can be realized in this setting as we now explain.
Since the Higgs curve is such that $\Sigma_H\cdot(E_1-E_2)=0$, the restriction of the 
$U(1)_{B-L}$ line bundle $L$ to $\Sigma_H$ has degree zero. We can then choose
\beq
L_{\Sigma_H}\ =\ \co_{\Sigma_H}(p_1-p_2)^{1/2} \ ,
\eeq
where $p_1$ and $p_2$ are two independent degree one divisors on $\Sigma_H$
(see the appendix).
We also know that under $SO(10)\supset SU(3)\times SU(2)_R \times SU(2)_L \times U(1)_{B-L}$ 
the $\r{10}$ has a branching 
\beq
\r{10}= (\r1,\r2,\r2,0) \ +\ (\br{3},\r1,\r1,2) \ +\ (\r3, \r1,\r1,-2) \, 
\eeq
so that the $B{\rm -}L$ flux will not affect the doublets that are neutral.
To compute the multiplicities according to
(\ref{multqq}) we still need to specify the bundle $L^\prime$. 
A simple option is to choose a trivial line bundle. In this way we obtain 
\beq
N(\r1,\r2,\r2,0) \ =\ h^0(\Sigma_{H}, \co_{\Sigma_H} )\ = \ 1 \ ,
\eeq
because on a curve of genus one the only holomorphic sections of the trivial bundle are the constant functions.
The upshot is that there is a massless Higgs doublet. Recalling that $\Sigma_H$ has trivial canonical line 
bundle, for the triplet we instead find
\beq
N(\br{3},\r1,\r1,2)  =  \ h^0(\Sigma_H, \co_{\Sigma_H}(p_1-p_2))  =  0 \ ,
\label{tripdeg} 
\eeq
because the divisor $(p_1-p_2)$ is not effective. Using this result together
with the index theorem (\ref{indexqq}), and the fact that $c_1(\co_{\Sigma_H}(p_1-p_2))=0$, 
it also follows that
\beq 
N(\r3,\r1,\r1,-2) = h^0(\Sigma_H, \co_{\Sigma_H}(p_1-p_2)^{-1}) = 0 \ .
\label{tripdeg2} 
\eeq
We thus obtain a minimal set of massless Higgs doublets.

We now consider the effect of the $U(1)_{B-L}$ flux on the $\r{16}$s living on the
two matter curves $\Sigma_1$ and $\Sigma_2$. To begin, we recall the well known branching of the
$\r{16}$ under $SO(10) \supset SU(3)\times SU(2)_R \times SU(2)_L \times U(1)_{B-L}$
\beq
\r{16} =  (\r1,\r2,\r1,3) +(\r3,\r1,\r2,1) +(\br{3},\r2,\r1,-1) + (\r1,\r1,\r2,-3) \ . 
\label{br16}
\eeq
Substituting  the bundle data of table \ref{so10} in (\ref{multqq}) one finds for the matter fields 
from  curve $\Sigma_1$ (with $q^\prime=1$) the multiplicities
\beqa
N(L_R) & = & N(\r1,\r2,\r1,3)  \hspace*{3mm} =  \
h^0(\Sigma _1, \co_{\Sigma_1}(-1+\frac {3}{2} +\frac {3}{2}))\ =\ 3 \ ,\nonumber \\
N(Q_L) & = & N(\r3,\r1,\r2,1)  \hspace*{3mm} = \  
h^0(\Sigma _1, \co_{\Sigma_1}(-1+\frac {1}{2} +\frac {3}{2}))\ =\ 2  \ ,
\label{mult16uno} \\
N(Q_R) & = & N(\br{3},\r2,\r1,-1) =  \ h^0(\Sigma _1, \co_{\Sigma_1}(-1-\frac {1}{2} +\frac {3}{2}))\ =\ 1 
\ ,\nonumber \\ 
N(L_L)& = & N(\r1,\r1,\r2,-3)  =  \ h^0(\Sigma_1, \co_{\Sigma_1}(-1-\frac {3}{2} +\frac {3}{2}))\ =\ 0 \ . \nonumber
\eeqa
For the second curve $\Sigma_2$ one instead obtains
\beqa
N(L_R) & = & N(\r1,\r2,\r1,3)  \hspace*{3mm} =  \
h^0(\Sigma_2, \co_{\Sigma_2}(-1-\frac {3}{2} +\frac {3}{2}))\ =\ 0 \ ,\nonumber \\
N(Q_L) & = & N(\r3,\r1,\r2,1)  \hspace*{3mm} = \  
h^0(\Sigma_2, \co_{\Sigma_2}(-1-\frac {1}{2} +\frac {3}{2}))\ =\ 1  \ , 
\label{mult16dos} \\
N(Q_R) & = & N(\br{3},\r2,\r1,-1) =  \ h^0(\Sigma_2, \co_{\Sigma_2}(-1+\frac {1}{2} +\frac {3}{2}))\ =\ 2 \ , 
\nonumber \\ 
N(L_L)& = & N(\r1,\r1,\r2,-3)  =  \ h^0(\Sigma_2, \co_{\Sigma_2}(-1+\frac {3}{2} +\frac {3}{2}))\ =\ 3 \ . \nonumber
\eeqa
Therefore, the matter content from each curve is
\beqa
\Sigma_1 & : &  3\times L_R + 2\times Q_L + 1\times  Q_R \ ,  \nonumber \\
\Sigma_2 & : & 3\times  L_L +  2\times Q_R + 1\times Q_L \ \label{dist16}.
\eeqa
Note that altogether there are three net generations but the fields have been redistributed and
there are no complete $\r{16}$s in any of the two curves.

The remarkable result is that now the rule that states that only fields coming from
different curves (i.e. $\Sigma_1 \times \Sigma_2 \times \Sigma_{H}$)
can have trilinear couplings leads to interesting textures. Indeed note 
that with the matter content summarized in (\ref{dist16}) the rule implies the Yukawa structure:
\beq
h_Q \ \sim \
\left(
\begin{array}{ccc}
x & x & 0 \\
x & x & 0 \\
0 & 0 & x
\end{array}
\right)
\ \ ;\ \
h_L \ \sim \
\left(
\begin{array}{ccc}
x & x & x \\
x & x & x\\
x & x & x
\end{array}
\right)
\label{textso10}
\eeq
where $x$ means something non-vanishing. We see that all fermions may now get a mass
without any self-interaction for the curves. This is important 
in itself but  as a byproduct we obtain three qualitative predictions:

\begin{itemize}
\item The third generation quarks mix  little with the first two generations.

\item First and second generations may have substantial (Cabbibo) mixing.

\item The mixing of leptons may generically be large.

\end{itemize}

\noindent
These three points are in good qualitative agreement with experiment.
(Actually, before the breaking of the left-right symmetry there is no mixing,
since the U-and D-quark mass matrices  are always proportional. The above statements apply once
the L-R symmetry has been broken and this proportionality ceases to be exact,
see comments at the end of the section).
Additional fermion hierarchies may result once one computes the
Yukawa couplings in terms of the values of internal wave 
functions at the intersection points as in 
eq.(\ref{Yukawas}). For example, the 3rd generation quarks 
will have larger Yukawas if the wave functions of $Q_L^3$,  $U_R^3$, and $D_R^3$
evaluated at some intersection point $p$ are larger than those of the corresponding
quarks  of the first two generations.
In particular, if the wave functions at this point satisfy
$\Psi_L^\tau(p)\Psi_R^\tau(p)\simeq \Psi_L^b(p)\Psi_R^b(p)\simeq
\Psi_L^t(p)\Psi_R^t(p)$, one would obtain relationships among
the 3rd generation Yukawas of the form $h_\tau \simeq h_b\simeq h_t$.

The gauge symmetry $SU(3)\times SU(2)_R\times SU(2)_L\times U(1)_{B-L}$
has to be further broken down to the SM group.
To this purpose we need to have vector-like right-handed doublets in our massless
spectrum. To obtain these fields we introduce a curve $\Sigma_\phi$
giving rise to multiplets $(\r{16}+\br{16})$ before flux is added. Then the flux must
split them so that only a vector-like pair of right-handed
doublets $[(\r1,\r2,\r1,3) + {\rm c.c.}]$ remains. This can be achieved
by considering a genus one curve where the restricted $U(1)_{B-L}$ bundle has degree zero again.
Some $B{\rm -}L$ flux must pierce the curve in order to split the representations and remove
all fields except the right-handed doublets. A possible curve can have a class 
\beq
  \Sigma_\phi = 3H -E_1-E_2 
\eeq
which  has $g=1$ and $\Sigma_\phi \cdot (E_1-E_2)=0$.
The restriction of the $B{\rm-}L$ flux onto $\Sigma_\phi$ can then again be taken of the form
\beq
                  L_{\Sigma_{\phi }}\  = \ \co_{\Sigma_\phi}(p_3-p_4)^{1/2} \ ,
\eeq
with $p_3$ and $p_4$ degree one divisors in $\Sigma_\phi$.
We now choose a non trivial restriction of the  $U(1)'$ bundle given by
\beq
     L^\prime_{\Sigma_\phi }\  = \ \co_{\Sigma_\phi}(p_3-p_4)^{-3/2}  \ .
\eeq
To compute the multiplicities we need the decomposition of the $\r{16}$ shown in (\ref{br16}).
Inserting into (\ref{multqq}) we find
\beq
    N(\r1,\r2,\r1,3) \ = \ h^0(\Sigma_\phi , \co_{\Sigma_\phi}) \   = \ 1 \ .
\eeq
The rest of the multiplets have zero multiplicity  since the $(p_3-p_4)$ divisor
is not effective and has degree zero. Thus, the only remaining fields are right-handed doublets
transforming as $[(\r1,\r2,\r1,3) + (\r1,\r2,\r1,-3)]$ whose vev would break the symmetry down to the SM.

The final spectrum is that of the MSSM. However there are still the three
copies of exotics $[(\br{3},\r1,\r1,-4)^{exot}+ {\rm c.c}]$ coming from the adjoint of $SO(10)$.
In fact, after the symmetry is broken down to the SM those multiplets
generically get massive.  Indeed, note that in general there ia a coupling
$\r{45} \times \r{16}_\phi \times  \br{16}_\phi$ that gives rise to
\beq
(\br{3},\r1,\r1,-4)^{exot}
\times  (\r1,\r2,\r1,3) \times  (\r3,\r2, \r1, 1)
\eeq
The triplets   $(\r3,\r2,\r1,1)$  inside $\Sigma_\phi$ are massive KK states and may be exchanged 
so that  an effective operator of the form
\beq
            (\r1,\r2,\r1,3) \times  (\r1,\r2,\r1,-3) \times  (\br{3},\r1,\r1,-4)^{exot}\times
(\r3,\r1, \r1, 4)^{exot}
\eeq
is generated.
When the right-handed doublets get a vev the three  pairs of triplets
from  the adjoint will  dissappear from the massless spectrum.

It is not the purpose of this paper to give a full phenomenological study
of the models presented. Several comments are however in order:

\begin{itemize}

\item
 Dirac neutrino masses appear generically
with the same size as the rest of quarks and leptons
from the couplings $\r{16}_1\times \r{16}_2\times \r{10}_H$. On the other hand, one
possible way to implement the see-saw mechanism is the following.
Large masses for the right-handed neutrinos (which live in the $\Sigma_1$ curve)
may appear if there are couplings of the form $\r{16}_1\times \br{16}_\phi\times N_i$,
with $N_i$ three SM singlets
(see figure \ref{so10quiver}). Once the right-handed doublets inside
$\br{16}_\phi$ get a vev, the three right-handed neutrinos become
massive and the standard sea-saw mechanism will be at work.
In this connection note that
$\Sigma_\phi$ has positive intersection with $\Sigma_1$ so that the required
coupling is in principle allowed.

\item
In this model R-parity is automatic due to the $SO(10)$ structure which
forbids $\r{16}^3$ couplings. Hence there are no dim=4 baryon or lepton number
violating couplings. On the other hand baryon number violating dim=5
operators (like e.g. $QQQL$)  are suppressed.
Such operators appear from couplings of quarks and leptons to the massive
color triplets inside the $\r{10}$. However, due to the split structure of the multiplets
in the $\r{16}$s one can check that the baryon number dim=5 operators always involve
at least one third-generation quark. Thus, the amplitude has additional
Cabbibo suppression which makes the exchange of colored Higgsses harmless.

\item
The  quark mass structure
in (\ref{textso10}) before the breaking of the
$SU(2)_R\times SU(2)_L\times U(1)_{B-L}$ symmetry gives mass
matrices for U- and D-quarks which are proportional so that at
this level there is in fact no mixing. However, once further symmetry breaking
takes place the insertion of vevs for the right-handed doublets
in $\Sigma_\phi$ gives rise to mixings between the
right-handed D-quarks from the $\r{16}$s and the color
triplets in the $\r{10}$ so that the massless eigenstates
are mixed linear combinations. Then  the
U- and D-quark mass matrices will  cease to be
proportional and mixing will generically occur.

\end{itemize}

We finish with some comments about possible extensions/generalizations of
the model. There is some freedom in the choice of line
bundles in the $U(1)^\prime_i$ associated to each matter curve.
For example, instead of bundles  $L^\prime_{\Sigma_i} = \co_{\Sigma_i}(3)^{1/2}$ 
for both $\Sigma_1$ and $\Sigma_2$ we could have chosen  $L^\prime_{\Sigma_1}=\co_{\Sigma_1}(4)^{1/2}$ 
and $L^\prime_{\Sigma_2}=\co_{\Sigma_2}(2)^{1/2}$, or even 
$L^\prime_{\Sigma_1}=\co_{\Sigma_1}(5)^{1/2}$ and $L^\prime_{\Sigma_2}=\co_{\Sigma_2}(1)^{1/2}$. 
With these alternative choices one
also gets three net quark/lepton generations. However one can easily  check that with
asymmetric assignments the massless spectrum includes extra
vector-like multiplets beyond the standard three generations.
Also in these cases always some quark or lepton remains massless.

The case with {\it three generations is also somewhat special} in these $SO(10)$ GUT's.
Indeed, although in principle one can get
any number of generations, three is the minimal number for which
the spectrum does not contain additional vector-like matter fields.
Note also in this respect that the bigger the number of generations, the bigger the
$U(1)^\prime$ instanton numbers $n_1$ and $n_2$. A larger number of generations would  make more difficult
the eventual embedding of the local model into a complete global F-theory compactification.
This is because $U(1)$ backgrounds induce $\D 3$-brane charge and the latter is bounded in a compact model.

One can also consider adding fluxes along the $U(1)$ in the branching of
$SO(10)$ into $SU(5)\times U(1)_5$. Flipped $SU(5)$ models with this
kind of flux were considered in \cite{bhv2}. However, it is easy to check that
allowing for this $U(1)$  flux to go through the matter curves does not add
anything new to the problem of Yukawa couplings. This is obvious since
the gauge group $SU(5)$ remains unbroken and its representations
contain both left- and right-handed fermions unified.

\section{A  $\pmb{SU(5)}$  F-theory GUT model}

We now discuss the case of $SU(5)$ GUT's. 
Consider then a bulk 7-brane with gauge group $SU(5)$. 
In the simplest situations $\r{10}$s are obtained at curves where
the $SU(5)$ singularity is enhanced to $SO(10)$ whereas 
$\br{5}$s will appear at curves where the 
singularity in enhanced to $SU(6)$. At points at which 
these curves intersect the symmetry is further enhanced 
to $E_6$ or $SO(12)$.
This is  the case of the examples considered in \cite{bhv2}.
In particular we have the branchings
\beqa
\hspace*{-5mm}E_6 & \supset & SU(5)\times  U(1)_1 \times U(1)_2  \label{rep78} \\ 
\r{78}  & =  & {\rm Adjoints }\ + \
\big[(\r{10},-1,-3) + (\r{10},4,0) + (\r5,-3,3) +(\r1,5,3) \ +\ {\rm c.c.} \big]  \nonumber \\ 
\hspace*{-5mm} SO(12) & \supset &  SU(5)\times U(1)_3\times U(1)_4 \label{rep66} \\ 
\r{66}  & =&  {\rm Adjoints }\ + \ \big[(\r{10},4,0) + (\br{5},-2,2) + (\br{5},-2,-2) \ +\ {\rm c.c.}\big]
\nonumber                              
\eeqa
The first branching shows that two $\r{10}$s from different Riemann curves
can form a Yukawa coupling with some $\r5$. On the other hand, the second shows that
one of the $\r{10}$ curves (but not the other) 
 may couple also to pairs of $\br{5}$'s to provide for
D-quark and lepton Yukawas.  One can now think of constructing a model
with two $\r{10}$ curves and one matter $\br{5}$ curve and repeating  the 
reshuffling idea we used before but using now hypercharge flux through the two
curves associated to the $\r{10}$s. However, it is easy to realize that   
this cannot possibly work. The problem is that after 
turning on the hyperflux, as we will see below, the  three generations
of quarks and leptons from the $\r{10}$s split into two sets
with $\Sigma_1$ and $\Sigma_2$ containing 
respectively $(3\times E_R+2\times Q_L+1\times U_R)$ 
 and $(2\times U_R+1\times Q_L)$. Now, according to our observation above,
only one of the $\r{10}$ curves (in particular $\Sigma_2$) may couple
to $\br{5}$s to form Yukawas. That means that only one
D-quark generation and no leptons would be allowed  to get Yukawas. 
This shows that in an F-theory $SU(5)$ GUT in which matter resides 
on curves with minimal  rank enhancing the presence of flux 
through the matter curves cannot possibly solve by itself the
Yukawa coupling problem.

\begin{figure}[t]
\epsfysize=9cm
\begin{center}
\leavevmode
\epsffile{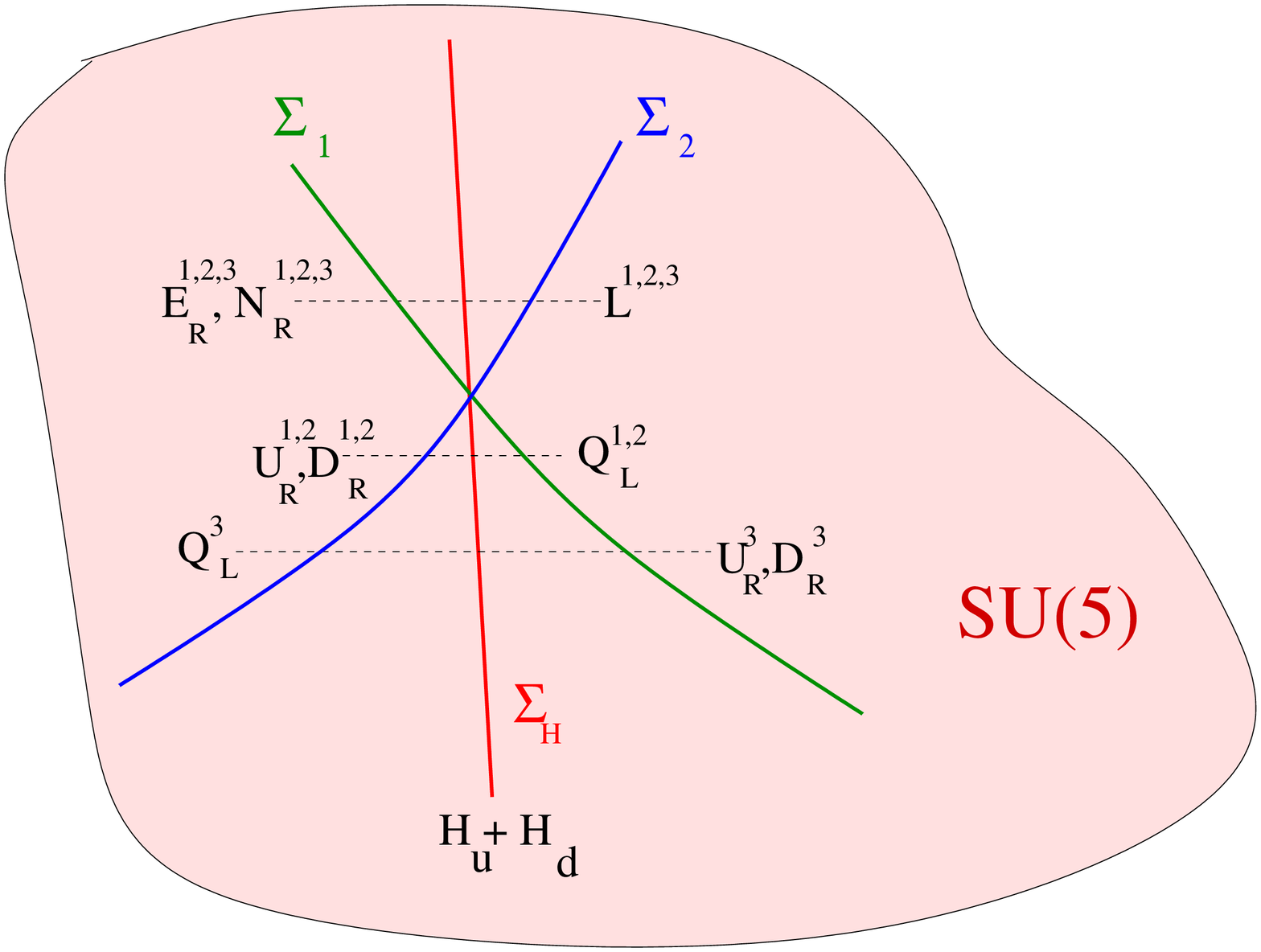}
\end{center}
\caption{\small Sketch of the structure of the hypercharge fluxed $SU(5)$ model.}
\label{su5quiver}
\end{figure}

On the other hand, the fact that $SO(10)$ contains $SU(5)$ as a subgroup
and that we were able to build a $SO(10)$ model with no Yukawa problem 
already tells us that it should be possible to construct $SU(5)$ GUT's
with the appropriate properties to solve it. Since
$SO(10)$ has rank higher in one unit this means that the adequate 
couplings should be possible if  in the matter curves  the $SU(5)$ 
singularity is enhanced in two units up  to $E_6$. This double enhancing in the
rank of the singularity is indeed possible, as remarked in \cite{hv1}.
In these matter curves both $\r{10}$s and $\br{5}$s can be present.
We know that a $\r{16}$ of $SO(10)$ contains these $SU(5)$ multiplets,
the relevant branching being 
\beqa
SO(10) & \supset & SU(5)\times  U(1)_5 \label{rep165} \\ 
\r{16}  & =  & (\r{10},-1) + (\br{5},3) + (\r1,-5) \ .    \nonumber
\label{desglose5}
\eeqa 
Then, the $E_7$ branching (\ref{rep133}) indicates that one can consider 
a pair of curves $\Sigma_1$ and $\Sigma_2$ each supporting a $\r{10}$,
a $\br{5}$, and a singlet. At the intersection of these curves the symmetry is enhanced to $E_7$,
just like in the $SO(10)$ GUT of the previous section.
As remarked in \cite{hv1}, in these doubly enhanced singularity curves 
there could be in principle additional matter fields from the further Higgssing of
$SO(10)$ down to $SU(5)$. In what follows we will
ignore this issue and simply examine what would be the effect on the Yukawas of non-vanishing 
fluxes through these matter curves.
An specific local realization of curves and bundles is summarized in  table \ref{tablasu5}. 
An sketch of the matter  curves involved  is depicted in figure \ref{su5quiver}.
The structure is quite analogous to the previous $SO(10)$ model. The main difference is that
we have an $SU(5)$ symmetry on $S$ which is broken down to the SM by hypercharge
fluxes. Furthermore, now there are  simultaneous fluxes along $U(1)^\prime\times U(1)_5$ that
break $G_{S^\prime}$. The corresponding line bundles are denoted $L^\prime$ and $\tilde L$.

\begin{table}[htb] \footnotesize
\renewcommand{\arraystretch}{1.25}
\begin{center}
\begin{tabular}{|c|c|c|c|c|c|}
\hline
 Curve  &  Class  &  $g_\Sigma $  & Multiplet & $ L_\Sigma $
 & $L^{\prime q^\prime}_\Sigma \otimes \tilde L^{q_5}_\Sigma$ \\
\hline\hline
$ \Sigma_1$  & $H-E_1-E_3$ &   0  &    ${\bf 10_1}$    &  $\co_{\Sigma_1}(1)^{1/5}$ & 
$\co_{\Sigma_1}(9)^{1/5}$  \\
          &       &        &     ${\bf {\overline 5}_1}$   &  
$\co_{\Sigma_1}(1)^{1/5}$ & $\co_{\Sigma_1}(3)^{1/5}$  \\
          &       &        &     ${\bf { 1}_1}$   &  $\co_{\Sigma_1}(1)^{1/5}$ &
 $\co_{\Sigma_1}(3)$  \\
\hline
$\Sigma_2$   & $H-E_2-E_4$ &   0   &   ${\bf 10_2  }$  &
 $\co_{\Sigma_2}(-1)^{1/5}$ & $\co_{\Sigma_2}(6)^{1/5}$  \\
         &       &        &     ${\bf {\overline 5}_2}$   & 
 $\co_{\Sigma_2}(-1)^{1/5}$ & $\co_{\Sigma_2}(12)^{1/5}$  \\
          &       &        &     ${\bf { 1}_2}$   & 
 $\co_{\Sigma_2}(-1)^{1/5}$ & $\co_{\Sigma_2}$  \\
\hline
$\Sigma_H$  &  $-K_S$  & 1  & ${\bf {\overline 5}_H}+ {\bf 5_H} $ &
  $\co_{\Sigma_H}(p_1-p_2)^{1/5} $ &  
$\co_{\Sigma_H}(p_1-p_2)^{-3/5}$      \\
\hline \end{tabular}
\end{center} \caption{\small  Curves and bundles of the $SU(5)$ model with
hypercharge flux and $L=\co _S(E_1-E_2)^{1/5}$.}
\label{tablasu5}
\end{table}
%

We start  from a surface $S$ with 7-branes corresponding to a
$SU(5)$ gauge group. Switching on magnetic flux through the hypercharge direction
breaks the GUT symmetry. The resulting hypercharge values of the matter fields
can be read off from the decompositions 
\beqa
SU(5) & \supset &  SU(3)\times  SU(2) \times U(1)_Y \label{b321} \\ 
\r{24}  & =  & {\rm Adjoints} + (\r3,\r2,-5) + (\br{3},\r2,5)    \nonumber \\
\r{10}  & =  & (\r3,\r2,1) + (\br{3},\r1,-4) + (\r1,\r1,6)   \nonumber \\
\br{5}  & =  & (\r1,\r2,-3) + (\br{3},\r1,2)    \nonumber 
\label{desglose6}
\eeqa
where the last entry denotes the hypercharge. As remarked in \cite{bhv2},  
with the choice 
\beq
L=\co_S(E_1-E_2)^{1/5}
\label{lsu5}
\eeq
the exotics $[(\r3,\r2,-5)+(\br{3},\r2,5)]$ are absent from the massless spectrum.

Let us now study the effect of the fluxes on the matter curves containing quarks and leptons.
We will assume  that  there are two genus zero curves
$\Sigma_1$ and $\Sigma_2$ in which the $SU(5)$ symmetry  is doubly enhanced
up to $E_6$. As we said, at each of these curves there are 
three types of multiplets, $\r{10}$,  $\br{5}$, and a singlet.
In the intersecting 7-branes wrapping $S_1'$ and $S_2'$ there are bundles along $U(1)^\prime$ and 
another $U(1)_5$. To make contact with the previous $SO(10)$ model it is important to realize that 
the $U(1)_{B-L}$ generator may be expressed in terms of hypercharge and $U(1)_5$ as follows
\beq
 Q_{B-L}\  = \ \frac  {1}{5} (2 Y -3 Q_5) \ .
\label{bmenosl}
\eeq
This relation helps to understand the data in  table \ref{tablasu5}.
Note that the first Chern classes of the restricted line bundles 
$L^{\prime q^\prime}_{\Sigma_1} \otimes \tilde L^{q_5}_{\Sigma_1}$ felt by the various
multiplets in $\Sigma_1$ are given by $-\frac{3}{10}q_5+\frac {3}{2}q'$, where $q^\prime=1$ and $q_5$ is 
the $U(1)_5$ charge in eq.(\ref{desglose5}). 
For $\Sigma_2$ the Chern classes are instead   $\frac{3}{10}q_5+\frac {3}{2}q'$.
Now, by virtue of eq.(\ref{bmenosl}), it follows that the total degree of
$L^Y_{\Sigma_1} \otimes L^{\prime q^\prime}_{\Sigma_1} \otimes \tilde L^{q_5}_{\Sigma_1}$ is equal
to $\frac12 q_{B_L} + \frac32 q^\prime$, which is precisely the total degree of 
$L^{q_{B-L}}_{\Sigma_1} \otimes L^{\prime q^\prime}_{\Sigma_1}$ in the $SO(10)$ model. 
For $\Sigma_2$ there is also perfect matching with the $SO(10)$ bundles in table \ref{so10}.
The important point is that along the matter curves $\Sigma_i$ there is actually 
$U(1)_{B-L}$ flux (rather than just hypercharge), with
$\Sigma_1$ and $\Sigma_2$ getting opposite fluxes.  Given this fact, everything is quite 
similar to the parent  $SO(10)$ model. In particular, it is easy to check that 
the fermions are redistributed as in the $SO(10)$ case. 

As an example let us see what happens with the fermions belonging to 
a $\r{10}$. As shown in the table, the restriction of the hypercharge bundle to
$\Sigma_1$ and $\Sigma_2$ are
\beqa
L_{\Sigma_1} &=& \co_{\Sigma_1}(1)^{1/5} \ , \nonumber \\[2mm]
L_{\Sigma_2} &=& \co_{\Sigma_2}(-1)^{1/5} \ , \label{instcondsu5} 
\eeqa
where we have used that $(E_1-E_2)\cdot \Sigma_1=1$ and $(E_1-E_2)\cdot \Sigma_2=-1$.
The spectrum from the  $\r{10}$ in $\Sigma_1$ can be deduced from the multiplicities
computed using the data in the table. Substituting  in (\ref{multqq}) yields
\beqa
N(E_R) & = & N(\r1,\r1,6)  \hspace*{3mm} =  \
h^0(\Sigma _1, \co_{\Sigma_1}(-1+\frac {6}{5} +\frac {9}{5}))\ =\ 3 \ ,\nonumber \\
N(Q_L) & = & N(\r3,\r2,1)  \hspace*{3mm} = \  
h^0(\Sigma _1, \co_{\Sigma_1}(-1+\frac {1}{5} +\frac {9}{5}))\ =\ 2  \ ,
\label{mult10uno} \\
N(U_R) & = & N(\br{3},\r1,-4) =  \ h^0(\Sigma _1, \co_{\Sigma_1}(-1-\frac {4}{5} +\frac {9}{5}))\ =\ 1 
\ .\nonumber 
\eeqa
Similarly, for the second curve $\Sigma_2$ one obtains
\beqa
N(E_R) & = & N(\r1,\r1,6)  \hspace*{3mm} =  \
h^0(\Sigma _2, \co_{\Sigma_2}(-1-\frac {6}{5} +\frac {6}{5}))\ =\ 0 \ ,\nonumber \\
N(Q_L) & = & N(\r3,\r2,1)  \hspace*{3mm} = \  
h^0(\Sigma _2, \co_{\Sigma_2}(-1-\frac {1}{5} +\frac {6}{5}))\ =\ 1  \ ,
\label{mult10dos} \\
N(U_R) & = & N(\br{3},\r1,-4) =  \ h^0(\Sigma _2, \co_{\Sigma_2}(-1+\frac {4}{5} +\frac {6}{5}))\ =\ 2 
\ .\nonumber 
\eeqa
The multiplets from both curves are then
\beqa
\Sigma_1 & : &  3\times E_R + 2\times Q_L + 1\times  U_R \ ,  \nonumber \\
\Sigma_2 & : & 2\times U_R + 1\times Q_L \ \label{dist10}.
\eeqa
This is the content of three $\r{10}$s of $SU(5)$ unequally distributed between
the two curves. It is easy to check that the full content in both curves is
\beqa
\Sigma_1 & : & 
[\, 3\times E_R \ +\ 2 \times Q_L \ +\ 1\times U_R \, ]_{10_1}\ +\ 
[\, 1\times D_R \, ]_{{\overline 5}_1} \ +\ [\, 3\times N_R \, ]_{1_1}
\nonumber \\
\Sigma_2 & : & 
[\, 1\times Q_L \ +\  2\times U_R \, ]_{10_2}\ +\  [\, 2\times D_R
\ +\ 3\times L_L \, ]_{{\overline 5}_2} 
\label{distfullsu5}
\eeqa
which corresponds to the same distribution we found in the $SO(10)$ model.

Assuming that there is a Higgs curve $\Sigma_{H}$ with triple intersection with 
$\Sigma_1$ and $\Sigma_2$ and looking at the distribution of fermions  one
observes that there is a structure for quark masses of the form
\beq
h_U \ \sim \
\left(
\begin{array}{ccc}
x & x & 0 \\
x & x & 0 \\
0 & 0 & x
\end{array}
\right)
\quad  ; \quad 
h_D \ \sim \
\left(
\begin{array}{ccc}
x & x & 0 \\
x & x & 0 \\
0 & 0 & x
\end{array}
\right) \ .
\label{utextsu51Q}
\eeq
For the lepton masses the pattern turns out to be
\beq
h_L \ \sim \
\left(
\begin{array}{ccc}
x & x & x \\
x & x & x\\
x & x & x
\end{array}
\right)
\quad ;\quad
h_N \ \sim \
\left(
\begin{array}{ccc}
x & x & x \\
x & x & x\\
x & x & x
\end{array}
\right) \ .
\label{utextsu51L}
\eeq
This is analogous to what we found for both U- and D-quarks in the
$SO(10)$ example, the main difference being that the mass matrices for
U- and D-quarks are now in general not strictly proportional since the
wave function of $U_R$ and $D_R$ fields may now differ and hence there will
be Cabbibo mixing. 
The same happens with charged leptons and Dirac neutrino masses which are no longer
proportional. Again the lesson is that a small mixing for the third generation is predicted,
whereas other mixings in the CKM (and neutrino mixing) are not
generically small.

The $SU(5)$ model may be considered in some sense as a variant 
of the parent $SO(10)$ model. However it has the advantage that the
symmetry breaking down to the SM does not require explicit
field theoretical Higgssing. It would be interesting to 
find an F-theoretical 7-brane recombination process connecting
both models. In any event it is clear  that in order to obtain
Yukawa couplings for all fermions the flux of a particular
$U(1)$, that of $U(1)_{B-L}$, is required.
In the $SO(10)$ model this flux acts in the bulk $S$ surface.
In $SU(5)$ the $U(1)_{B-L}$ flux goes through the matter curves, while in the
bulk only hypercharge flux is felt.

Let us make some comments about the phenomenological properties of these $SU(5)$ models.
As in  the $SO(10)$ case, right-handed neutrinos appear generically in the spectrum. 
However, in the present case possible Majorana right-handed neutrino masses are not forbidden by the 
bulk gauge interactions but only by the $U(1)_5$ and $U(1)^\prime$ symmetries whose gauge bosons  
are presumably massive via combination with RR-fields.  Under these conditions Majorana 
masses for right-handed neutrinos could appear induced by stringy instanton effects \cite{instantons}.
In this  model R-parity is automatic due to the absence of self-couplings of matter $\br5$-plets.
In addition baryon number violating dim=5 operators are suppressed just like in its parent $SO(10)$ model.

At this level the third generation quarks  have  no mixing 
with the first two generations. Experimentally that mixing is of order 
$10^{-2}{\rm -}10^{-3}$ so this is a good starting point. Eventually some corrections 
should  give rise to this mixing. A natural possibility is again 
instanton induced couplings. Notice in this respect that
such mixing is forbidden to leading order by the $U(1)'$ symmetries 
under which the fermion curves are charged. Those $U(1)$'s are generically 
anomalous and their gauge boson massive so that instanton effects of the
type considered in \cite{instantons} could  give rise to the required
non-vanishing but small third generation mixing. Instanton effects inducing 
 non-perturbative corrections to Yukawa couplings 
in perturbative Type II orientifold models have been 
recently considered in \cite{yukor}.

\section{Final comments and outlook}

Yukawa couplings in F-theory GUT's come from the intersection of three
different matter curves. At  first sight this fact makes the structure of Yukawa couplings
unrealistic, unless one assumes that the involved matter curves have a self-intersecting
structure. In this paper we have remarked that a slight generalization of the
conditions  on the $U(1)_Y$/$U(1)_{B-L}$ fluxes breaking the GUT symmetry
solves this problem in a natural and attractive way. When we have 
two matter curves $\Sigma_1$ and $\Sigma_2$  giving rise to the same GUT representation, instead of 
requesting that  these fluxes vanish in each individual matter curve
it is enough to impose that the fluxes, particularly $U(1)_{B-L}$ fluxes,  cancel on average
(i.e.$\int_{\Sigma_1}F_{U(1)_{B-L}}+\int_{\Sigma_2}F_{U(1)_{B-L}}=0$). 
Under these conditions the SM fermions redistribute asymmetrically on the two matter
curves so that Yukawa couplings for all quarks and leptons are allowed.
This structure predicts suppressed CKM mixing for  the third quark generation
and unsuppressed mixing for the rest of the generations and also for leptons.
This prediction is in good qualitative agreement with experiment.

We have constructed specific local F-theory $SO(10)$ and $SU(5)$ 
configurations with this structure.
It would be interesting to further develop the phenomenology of these models.
In particular it would be interesting to explore whether one can find
configurations with good quantitative agreement with observed 
quark/lepton masses and mixings. These models have the low-energy
spectrum of the MSSM. The structure of  SUSY-breaking soft terms
under the assumption of modulus dominance has recently been analysed
in \cite{aci}, where compatibility with low-energy
experimental constraints and radiative electroweak symmetry 
breaking was shown. The possibility that  SUSY is induced by a particular
variety of gauge mediation has also been recently studied in
\cite{mss1,hv1}. It would be interesting to see whether the
splitting of matter fields on two different curves $\Sigma_1$ and $\Sigma_2$, 
as explored in the present paper, has a bearing on these and 
other phenomenological aspects of F-theory GUT's. 
Let us finally comment that the smallness of the mixing 
of the third quark generation with the first two families  could 
perhaps be a first hint, together with gauge coupling unification,  of 
an underlying F-theory grand unification.

\vspace*{2cm}

{\bf Acknowledgments}\\
We thank  F. Marchesano,  A. Uranga, and specially
S. Theisen, for useful discussions.
A.F. acknowledges a research grant No. PI-03-007127-2008 from CDCH-UCV. 
This work has been supported by the European
Commission under RTN European Programs MRTN-CT-2004-503369,
MRTN-CT-2004-005105, by the CICYT (Spain) under project
FPA2006-01105, the Comunidad de Madrid under project HEPHACOS
P-ESP-00346 and the Ingenio 2010 CONSOLIDER program CPAN.

\clearpage

\appendix
\section{Brief compendium on del Pezzo surfaces and other mathematical results}

The first step to construct an F-theory GUT is to specify the surface $S$
wrapped by the 7-branes. In this work we mostly consider $S=dP_8$.  
The del Pezzo surfaces $dP_N$, $N=1, \cdots, 8$, are defined as the blowup of $\pp^2$ at $N$ points.  
The canonical class of $dP_N$ is
\beq
K_S=-c_1(S)=-3H + \sum_{i=1}^N E_i  \ ,
\label{dpdata1}
\eeq
where $H$ is the hyperplane class and the $E_i$ are exceptional divisors. We will
also need the intersections
\beq
H\cdot H=1 \quad ; \quad H\cdot E_i=0 \quad ; \quad E_i\cdot E_j=-\delta_{ij} \ .
\label{dpdata2}
\eeq
The generators $\a_i=E_i - E_{i+1}$, $i=1, \cdots, N-1$, and $\a_N=H-E_1-E_2-E_3$, 
have intersection products equal to minus the Cartan matrix of the Lie algebra $E_N$
and can be regarded as simple roots. 

An important result is that the admissible line bundles
on $dP_N$ are in one to one correspondence with roots of the $E_N$ algebra \cite{bhv1, bhv2}.
For example, the line bundle of the $U(1)$ flux that breaks the GUT group can be chosen
to be $L=\co_S(\a_i)^{1/n}$. Fractional bundles are allowed as long as the physically
relevant powers are integers. 

After selecting the surface $S$ one has to define the matter curves $\Sigma_i$ that
must be divisors of $S$. Since $S$ has complex dimension two, the $\Sigma_i$ are Riemann
surfaces and their genus is given by
\beq
2g_i -2 = \Sigma_i\cdot(\Sigma_i + K_S) \ .
\label{genuss}
\eeq
For instance, for $S=dP_N$, taking $\Sigma_i=E_i$ yields $g_i=0$, as it should because each $E_i$
is a $\pp^1$. We will only consider curves of genus zero or one.   

To evaluate the degeneracies of charged multiplets living on a matter curve $\Sigma$ we need to
know 
$h^0(\Sigma, K_{\Sigma}^{1/2}\otimes L^q_{\Sigma} \otimes L^{\prime q^\prime}_{\Sigma})=
{\rm dim}_{\complexp} 
H_{\bar \partial}^0(\Sigma, K_{\Sigma}^{1/2}\otimes L^q_{\Sigma} \otimes L^{\prime q^\prime}_{\Sigma})$.
The product of line bundles is another line bundle of degree equal to the degree of
its associated divisor. A useful result is that when $\Sigma$ has genus zero $h^0(\Sigma, \co_\Sigma(d))$
vanishes when the degree $d$ is negative and it is equal to $(d+1)$ for $d \geq 0$.

The restricted bundle $L_\Sigma$ is determined by the line bundle $L$ on $S$ that is introduced
from the start to break the GUT group $G_S$. Since $L$ has a corresponding divisor $D$ on $S$,
$L_\Sigma$ will be a line bundle of degree equal to $\Sigma \cdot D$. On the other hand, the
form of the line bundle $L^\prime$ on the intersecting surface $S^\prime$ need not be given explicitly
because the gauge theory on $S^\prime$ is completely decoupled. One only has to specify the restriction
$L^\prime_\Sigma$. 

When $\Sigma$ has genus zero the restricted line bundles can be characterized by their degrees.
However, when $\Sigma$ has genus one it is necessary to give the associated divisor that
might be effective or not. A divisor $\cd$ of $\Sigma$ can be written as a formal linear combination
of irreducible codimension one hypersurfaces. Since $\Sigma$ is a curve, these hypersurfaces
are points $p_m$ and $\cd = \sum_m a_m p_m$. The divisor is effective if $a_m > 0$, $\forall m$. 
The degree of $\cd$ is equal to $\sum_m a_m$.


\clearpage


{\small

\begin{thebibliography}{99}


\bibitem{aiqu}
  G.~Aldazabal, L.~E.~Ib\'a\~nez, F.~Quevedo and A.~M.~Uranga,
  ``D-branes at singularities: A bottom-up approach to the string  embedding of
  the standard model,''
  JHEP {\bf 0008}, 002 (2000)
  [arXiv:hep-th/0005067].


\bibitem{bjl}
  D.~Berenstein, V.~Jejjala and R.~G.~Leigh,
  ``The standard model on a D-brane,''
  Phys.\ Rev.\ Lett.\  {\bf 88} (2002) 071602
  [arXiv:hep-ph/0105042].

\bibitem{cgqu}
  J.~F.~G.~Cascales, M.~P.~Garcia del Moral, F.~Quevedo and A.~M.~Uranga,
  ``Realistic D-brane models on warped throats: Fluxes, hierarchies and  moduli
  stabilization,''
  JHEP {\bf 0402} (2004) 031
  [arXiv:hep-th/0312051].

\bibitem{vw}
  H.~Verlinde and M.~Wijnholt,
  ``Building the standard model on a D3-brane,''
  JHEP {\bf 0701} (2007) 106
  [arXiv:hep-th/0508089]\\
  D.~Malyshev and H.~Verlinde,
  ``D-branes at Singularities and String Phenomenology,''
  Nucl.\ Phys.\ Proc.\ Suppl.\  {\bf 171} (2007) 139
  [arXiv:0711.2451 [hep-th]].


\bibitem{cmq}
J.~P.~Conlon, A.~Maharana and F.~Quevedo, \ \ 
``Towards Realistic String Vacua,'' arXiv:0810.5660 [hep-th]. 



\bibitem{ftheory}
  C.~Vafa,
  ``Evidence for F-Theory,''
  Nucl.\ Phys.\  B {\bf 469} (1996) 403
  [arXiv:hep-th/9602022]\\
  D.~R.~Morrison and C.~Vafa,
  ``Compactifications of F-Theory on Calabi--Yau Threefolds  I,''
  Nucl.\ Phys.\  B {\bf 473} (1996) 74
  [arXiv:hep-th/9602114];
``Compactifications of F-Theory on Calabi--Yau Threefolds II,''
  Nucl.\ Phys.\  B {\bf 476} (1996) 437
  [arXiv:hep-th/9603161];\\
For a recent review see  F.~Denef,
  ``Les Houches Lectures on Constructing String Vacua,''
  arXiv:0803.1194 [hep-th].


\bibitem{dw}
  R.~Donagi and M.~Wijnholt,
  ``Model Building with F-Theory,''
  arXiv:0802.2969 [hep-th].


\bibitem{bhv1}
  C.~Beasley, J.~J.~Heckman and C.~Vafa,
  ``GUTs and Exceptional Branes in F-theory - I,''
  arXiv:0802.3391 [hep-th].

\bibitem{bhv2}
  C.~Beasley, J.~J.~Heckman and C.~Vafa,
  ``GUTs and Exceptional Branes in F-theory - II: Experimental Predictions,''
  arXiv:0806.0102 [hep-th].


\bibitem{gcu}
S.~Dimopoulos, S.~Raby and F.~Wilczek,
``Supersymmetry And The Scale Of Unification,''
Phys.\ Rev.\ D {\bf 24} (1981) 1681;\\
L.~E.~Ib\'a\~nez and G.~G.~Ross,
``Low-Energy Predictions In Supersymmetric Grand Unified Theories,''
Phys.\ Lett.\ B {\bf 105} (1981) 439;\\
S.~Dimopoulos and H.~Georgi,
``Softly Broken Supersymmetry And SU(5),''
Nucl.\ Phys.\ B {\bf 193} (1981) 150.


\bibitem{aci}
  L.~Aparicio, D.~G.~Cerde\~no and L.~E.~Ib\'a\~nez,
  ``Modulus-dominated SUSY-breaking soft terms in F-theory and their test at
  LHC,''
  JHEP {\bf 0807}, 099 (2008)
  [arXiv:0805.2943 [hep-ph]].


\bibitem{tw}
  R.~Tatar and T.~Watari,
  ``GUT Relations from String Theory Compactifications,''
  arXiv:0806.0634 [hep-th].

\bibitem{httwy}
  H.~Hayashi, R.~Tatar, Y.~Toda, T.~Watari and M.~Yamazaki,
  ``New Aspects of Heterotic--F Theory Duality,''
  Nucl.\ Phys.\  B {\bf 806}, 224 (2009)
  [arXiv:0805.1057 [hep-th]].

\bibitem{hmssv}
  J.~J.~Heckman, J.~Marsano, N.~Saulina, S.~Schafer-Nameki and C.~Vafa,
  ``Instantons and SUSY breaking in F-theory,''
  arXiv:0808.1286 [hep-th].

\bibitem{hv1}
  J.~J.~Heckman and C.~Vafa,
  ``F-theory, GUTs, and the Weak Scale,''
  arXiv:0809.1098 [hep-th].

\bibitem{hv2}
  J.~J.~Heckman and C.~Vafa,
  ``From F-theory GUTs to the LHC,''
  arXiv:0809.3452 [hep-ph].

\bibitem{mss1}
  J.~Marsano, N.~Saulina and S.~Schafer-Nameki,
  ``Gauge Mediation in F-Theory GUT Models,''
  arXiv:0808.1571 [hep-th].

\bibitem{mss2}
  J.~Marsano, N.~Saulina and S.~Schafer-Nameki,
  ``An Instanton Toolbox for F-Theory Model Building,''
  arXiv:0808.2450 [hep-th].

\bibitem{w}
  M.~Wijnholt,
  ``F-Theory, GUTs and Chiral Matter,''
  arXiv:0809.3878 [hep-th].

\bibitem{dw2}
  R.~Donagi and M.~Wijnholt,
  ``Breaking GUT Groups in F-Theory,''
  arXiv:0808.2223 [hep-th].


\bibitem{imr}
  L.~E.~Ibanez, F.~Marchesano and R.~Rabadan,
  ``Getting just the standard model at intersecting branes,''
  JHEP {\bf 0111} (2001) 002
  [arXiv:hep-th/0105155].


\bibitem{bmmvw}
  M.~Buican, D.~Malyshev, D.~R.~Morrison, H.~Verlinde and M.~Wijnholt,
  ``D-branes at singularities, compactification, and hypercharge,''
  JHEP {\bf 0701} (2007) 107
  [arXiv:hep-th/0610007].


\bibitem{instantons}
  R.~Blumenhagen, M.~Cvetic and T.~Weigand,
  ``Spacetime instanton corrections in 4D string vacua - the seesaw mechanism
  for D-brane models,''
  Nucl.\ Phys.\  B {\bf 771}, 113 (2007)
  [arXiv:hep-th/0609191]\\
  L.E. Ib\'a\~nez  and A.~M.~Uranga,
  ``Neutrino Majorana masses from string theory instanton effects,''
  JHEP {\bf 0703} (2007) 052
  [arXiv:hep-th/0609213].


\bibitem{yukor}
  R.~Blumenhagen, M.~Cvetic, D.~Lust, R.~Richter and T.~Weigand,
  ``Non-perturbative Yukawa Couplings from String Instantons,''  \ 
  Phys.\ Rev.\ Lett.\ \,  {\bf 100} (2008) 061602
  [arXiv:0707.1871 [hep-th]]\\
 L.~E.~Ib\'a\~nez and A.~M.~Uranga,
  ``Instanton Induced Open String Superpotentials and Branes at
  Singularities,''
  JHEP {\bf 0802} (2008) 103
  [arXiv:0711.1316 [hep-th]]\\
  L.~E.~Ib\'a\~nez and R.~Richter,
  ``Stringy Instantons and Yukawa Couplings in MSSM-like Orientifold Models,''
  arXiv:0811.1583 [hep-th].





\end{thebibliography}
}
\end{document}